# Compressive Sensing Based Situational Awareness and Sensor Placement for DC Microgrids with Relatively Fixed Operation Patterns

Shutang You[1], Yilu Liu[1,2]
1. The University of Tennessee at Knoxville, Knoxville, TN, USA
2. Oak Ridge National Laboratory, Oak Ridge, TN, USA
Email: syou3@utk.edu; liu@utk.edu

***Abstract*— This paper proposes a state estimation and sensor placement method for DC microgrids that has relatively fixed operation patterns. It is developed based on compressive sensing theory. Formulations of various types of measurements and components are developed under the proposed framework. A measurement placing strategy to minimize the coherence of the measurement matrix and thus increase estimation accuracy is presented. Simulation results show that the proposed state estimation and sensor placing approach can effectively reduce the number of sensors to achieve a certain level of estimation accuracy.**

## I. Introduction

Microgrid is important for integrating renewables and distributed resources for improving power grid reliability, and reducing cost and environmental impact. Due to these increasing inverter-based generation, energy storages, and loads, DC microgrid is becoming a popular option for microgrid configuration and its control problems have been intensively studied recently. Distributed generation such as wind, PV, and fuel cells are based on DC or have internal DC links. Additionally, to improve better controllability and efficiency, loads are also becoming more based on DC power [1, 2]. For example, the various-speed motors, electronic devices, and heating. Distributed sources such as batteries and super capacitors, are also categorized as DC devices inherently.

For security, large-scale power systems are equipped with dedicate situation awareness and control systems, which consist of sensors, advanced analysis algorithms, and multiple actuators [3]. The increase of renewables adds more difficulties as well as value of system situation awareness [4, 5]. Similar to large power systems, it is important to understand DC microgrid operation situation for various control and protection purposes [6]. The variation and uncertainty of distributed resources brings challenges for understanding the real-time microgrid operation situation [7]. As the microgrid provides flexibility for integrating or shut down a substantial amount of distributed resources without having to report to control centers as in large power grids, the microgrid is more vulnerable to drastic voltage changes and power flow fluctuations [8]. Moreover, many distributed generators and loads do not have meters installed at the point of connection, which makes it difficult for situational awareness compared with conventional large power grids.

There are some pioneer studies that apply compressive sensing has in power systems. For example, Ref. [9, 10] applied compressive sensing to obtain higher resolution when observing harmonics and interharmonics. Ref. [11] used compressive sensing to detect the fault location in distribution systems. Ref. [12] studied new methods based on the idea of compressive sensing to correct spare error from nonlinear measurements. Ref. [13] and [14] proposed a new method based on compressive sensing for topology identification. Ref. [15] studied the strong coupling of voltage phasors in distribution systems, which allows the application of compressive sensing in distribution system state estimation. In [16, 17], compressive sensing was applied to reconstruct missing and bad synchrophasor data and reduce communication bandwidth requirements in wide-area measurement systems. Ref. [18] applied compressive sensing as a signal denoising method in power line communications. In [19, 20], compressive sensing was applied to power line outages identification by considering the power network as single graph. The issue of high coherence in the sensing matrices was tackled by matrix decomposition [21].

This paper focuses on the situational awareness of DC microgrids with relatively fixed operation patterns, which means that the system power flow pattern is relative fixed at one or several states most of the time and it does not deviate much from these states due to additional loads or generators. The deviation from these fixed operation patterns are some random connections of additional generation or loads on just a few buses. As these random factors happen in a sparse manner, the situational awareness of such DC microgrids can be achieved using fewer sensors and probably achieve higher accuracy provided the knowledge of these fixed operation patterns.

The main contribution of this paper is to propose a state estimation method based on compressive sensing (CS) to improve DC microgrid situation awareness. Under the compressive sensing framework, the formulations of various measurements and components including voltage, current, and power pseudo and real sensors are developed for DC grid state estimation using fewer meters. To improve estimation performance, a meter placing method is presented to minimize the coherence of the measurement matrix. The proposed state estimation and meter placement methods are tested on the standard DC power flow representation of the IEEE 9 bus and 118 bus systems.

This work is primarily supported by NSF under Award Number ECCS-1931975. This work made use of Engineering Research Center shared facilities supported by the Engineering Research Center Program of the National Science Foundation and the Department of Energy under NSF Award Number EEC-1041877 and the CURENT Industry Partnership Program.

## II. Situational Awareness for DC Microgrids with Fixed Operation Patterns based on Compressive Sensing

### *A. State Estimation based on Compressive Sensing for DC Microgrid with Fixed Operation Patterns*

Compressive sensing represents a category of methods to reconstruct signals with sparsity properties using far less samplings required by the Shannon-Nyquist sampling theory. It has been applied in many areas since it was proposed by Candès and Tao in 2006 [22]. For example, Ref. [23] studied the application of compressive sensing in improving the frequency resolution of power quality monitoring without increasing measurement time significantly. Ref. [24] used compressive sensing to compress PMU data and reconstruct missing data. Ref. [25] applied compressive sensing to reduce feedback overhead in communication systems.

The essence of compressive sensing is to take advantage of the sparsity of a signal to find a unique spare solution of an underdetermined linear system. A pre-requisite of compressed sensing is that the original signal $\boldsymbol{x}$ has sparsity in some domain, which can be represented by a transformation matrix basis $\boldsymbol{\varphi} \in \mathbf{R}^{N \times N}$ and a sparse vector $\boldsymbol{\alpha}$:

$$\boldsymbol{x} = \boldsymbol{\varphi\alpha} \tag{1}$$

$\boldsymbol{\alpha}$ is a sparse vector as it has far less non-zero elements than its dimensions. The aim of compressed sensing is to reconstruct $\boldsymbol{x}$ based on a measurement vector $\boldsymbol{y}$ [26]:

$$\boldsymbol{y} = \boldsymbol{\sigma x} = \boldsymbol{\sigma\varphi\alpha} \tag{2}$$

where $\boldsymbol{\sigma}$ is the measurement matrix. The sparsity index $S$, denoting the sparsity of $\boldsymbol{x}$, is the number of non-zeros elements in $\boldsymbol{\alpha}$. Let $N$ denote the dimension of the measurement vector $\boldsymbol{y}$. $M$ denotes the dimension of the original signal $\boldsymbol{x}$. As $N$ is much smaller than $M$, the reconstruction of $\boldsymbol{x}$ is realized by solving a $l_0$ basis pursuit optimization problem.

$$\widehat{\boldsymbol{\alpha}} = \arg\min \|\boldsymbol{\alpha}\|_0 \tag{3}$$

s.t.

$$\boldsymbol{y} = \boldsymbol{\sigma\varphi\alpha} \tag{4}$$

or an $l_1$ basis pursuit denoising problem

$$\widehat{\boldsymbol{\alpha}} = \arg\min \|\boldsymbol{\alpha}\|_1 \tag{5}$$

s.t.

$$\|\boldsymbol{y} - \boldsymbol{\sigma\varphi\alpha}\|_2 < \boldsymbol{\epsilon} \tag{6}$$

Through solving the optimization problem, $\boldsymbol{x}$ can be reconstructed provided $M \geq C\mu^2(\boldsymbol{\sigma}, \boldsymbol{\varphi}) S \log N$. $\mu(\boldsymbol{\sigma}, \boldsymbol{\varphi})$ is the coherence of the column pairs in $\boldsymbol{\sigma}$ and $\boldsymbol{\varphi}$. $C$ is a constant.

The aim of microgrid state estimation is to provide a reliable result of the microgrid state based on all available measurements. The estimation formulation can be over-determined or under-determined equations depending on the measurement availability. In DC microgrids with relatively fixed operation patterns, deviations of components from these fixed patterns appear in a sparse manner in the time domain, which means the percentage of deviated components is small for most of the time. For example, some loads occasionally deviate from their normal values, and they seldom deviate simultaneously. It would be uneconomic to install a meter for each bus. As compressive sensing casts an insight that sparse signals can be accurately reconstructed using less measurements, it can be borrowed by situational awareness for such DC microgrids based on under-determined equations. Through sparse sensing, deviations of state variables of these distributed resources or loads from their values in fixed patterns are treated as sparse signals. Then, the system state variables could be fully reconstructed. Below are the formulations of various components under the state estimation framework based on compressive sensing.

*1) Voltage meters and voltage sources:* The basic formulation is a microgrid that contains generators and some known constant-resistance loads. The voltage measurements are used to estimate generator current and load current deviations. The network impedance matrix $\mathbf{Z}$ directly serves a projection between the sparse current injection deviation vector and the measured voltage deviation.

$$\begin{bmatrix} V_1 \\ V_2 \\ \cdot \\ V_N \end{bmatrix} = \begin{bmatrix} Z_{11} & Z_{12} & \cdots & Z_{1M} \\ Z_{21} & Z_{22} & \cdots & Z_{2M} \\ \cdot & \cdot & \cdots & \cdot \\ Z_{N1} & Z_{N2} & \cdots & Z_{NM} \end{bmatrix} \begin{bmatrix} I_1 \\ I_2 \\ \cdot \\ I_M \end{bmatrix} \tag{7}$$

$V_i$ is the voltage deviation at the bus $i$. $I_j$ is the generator or load current deviation at bus $j$. $Z_{ij}$ is the entry of the network resistance matrix $\mathbf{Z}$. When constant-resistance load exists, $\mathbf{Z}$ could be the modified network impedance matrix that incorporates load resistance. Despite that $N < M$, indicating the equations are underdetermined, generator current injection deviations and load current deviations can be used to reconstruct by solving the problems.

If a voltage regulated source is present, it can be treated as a voltage meter with an unknown current injection deviation, equivalent to adding a variable and a measurement.

*2) Current meters and current sources:* When current injection measurements or current regulated sources are present, the above equation can be updated by an offset. This scenario also covers the situation of constant current load deviations.

$$\begin{bmatrix} V_1' \\ V_2' \\ \cdot \\ V_N' \end{bmatrix} = \begin{bmatrix} Z_{11} & Z_{12} & \cdots & Z_{1M} \\ Z_{21} & Z_{22} & \cdots & Z_{2M} \\ \cdot & \cdot & \cdots & \cdot \\ Z_{N1} & Z_{N2} & \cdots & Z_{NM} \end{bmatrix} \begin{bmatrix} I_1 \\ I_2 \\ \cdot \\ I_M \end{bmatrix} + \begin{bmatrix} Z_{11} & Z_{12} & \cdots & Z_{1K} \\ Z_{21} & Z_{22} & \cdots & Z_{2K} \\ \cdot & \cdot & \cdots & \cdot \\ Z_{N1} & Z_{N1} & \cdots & Z_{NK} \end{bmatrix} \begin{bmatrix} I_{k1} \\ I_{k2} \\ \cdot \\ I_{kK} \end{bmatrix} \tag{8}$$

where $I_{ki}$ is the measured current deviation or the current deviation value of current-regulated sources.

*3) Constant power sources and loads:* Some generators in DC microgrid have constant power characteristics, such as MPPT control. As these components show negative impedance in small signal stability, estimating their current or voltage values is important to ensure stable operation. Since constant power sources and loads involves nonlinearity in state estimation, they require an iterative process based on the Jacobian matrix. The entries of Jacobian matrix $\boldsymbol{H}$ can be represented by

$$H_{ij} = \left[\frac{\partial P_i}{\partial I_j}\right] \tag{9}$$

where $P_i$ and $I_j$ are the power and current deviations of the source (load), respectively. The entry values can be calculated as

$$\begin{cases} \frac{\partial P_i}{\partial I_i} = 2Z_{ii}I_i + \sum_{i \neq j} Z_{ij}I_j \\ \frac{\partial P_i}{\partial I_j} = Z_{ij}I_i \qquad\qquad i \neq j \end{cases} \tag{10}$$

*B. Sensor Placement for Low Coherence of the Measurement Matrix*

In compressive sensing, a low coherence is desired in the measurement matrix so that fewer measurements will be needed to reconstruct system states. It has been shown that smaller off-diagonal elements' absolute values in the measurement matrix will lead to smaller coherence [27, 28]. In order to quantify the coherence, the Gram matrix is applied as $\boldsymbol{G} = \widetilde{\boldsymbol{D}}^T\widetilde{\boldsymbol{D}}$, where $\widetilde{\boldsymbol{D}}$ is the column-normalized version of $\boldsymbol{D}$, which equals $\boldsymbol{\sigma}$ as $\boldsymbol{\varphi} = \boldsymbol{I}_{MM}$. The ideal zero-coherence measurement matrix will have $\boldsymbol{G} = \boldsymbol{I}_{MM}$, where $\boldsymbol{I}_{MM}$ is the identity matrix. However, this will be impossible for the environment of using fewer measurements to estimate system states by compressive sensing. The idea here is to make the measurement matrix $\boldsymbol{G}$ as close as the identity matrix.

$$\widehat{\boldsymbol{G}} = \arg\min_{\boldsymbol{G}} \|\boldsymbol{G} - \boldsymbol{I}\|_{max}^2 \tag{11}$$

where $\|.\|_{max}$ is the matrix max norm. To minimize the max norm, a heuristic method to place the sensor in DC microgrid is shown in Fig. 1 with its major step explained as follows:

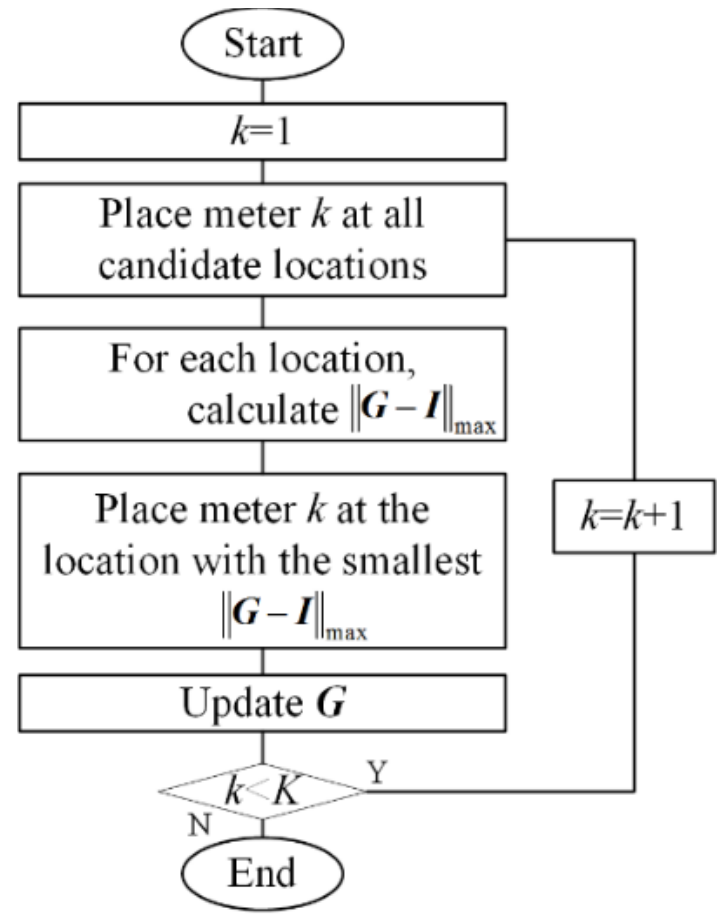

Fig. 1. Sensor placement procedures.

1. Calculate the $\|\boldsymbol{G} - \boldsymbol{I}\|_{max}^2$ when placing one measurement at each of the candidate locations.
2. Select the location that has the smallest max norm as the measurement location of the current sensor.
3. Update the measurement matrix.
4. Proceed the next measurement to archive the smallest max norm.

In DC microgrids, the bus impedance matrix could be ill-conditioned because of some short line sections. This ill-condition feature will increase the coherence of the deviation measurement matrix, thus requiring more meters to acquire the same level of estimation accuracy. In some situations, this feature makes it almost impossible to further reduce the number of sensors. The proposed meter placement method can search the best placement plan under this theoretical constraint.

In the real-world application, location of meters for voltage or other DC grid measurements prioritizes safety (e.g. protection) and fast response as highest. The proposed sensor placement method is to gain higher observability of the entire system, which is one step further to enable more advanced and selective protection schemes. If the system already has some existing sensors, which are most likely in practice, additional sensors can still be placed based on the proposed method: each additional sensor can be placed at the location that achieves the minimal Gram matrix.

## III. Case Studies

The DC representation of the IEEE 9 bus system and the IEEE 118 bus system are used to test the effectiveness of proposed state estimation and meter placement approaches.

*A: IEEE 9 Bus System*

The diagram of the IEEE 9 bus system is shown in Fig. 2. This system is converted to the DC microgrid model based on its standard DC power flow network representation.

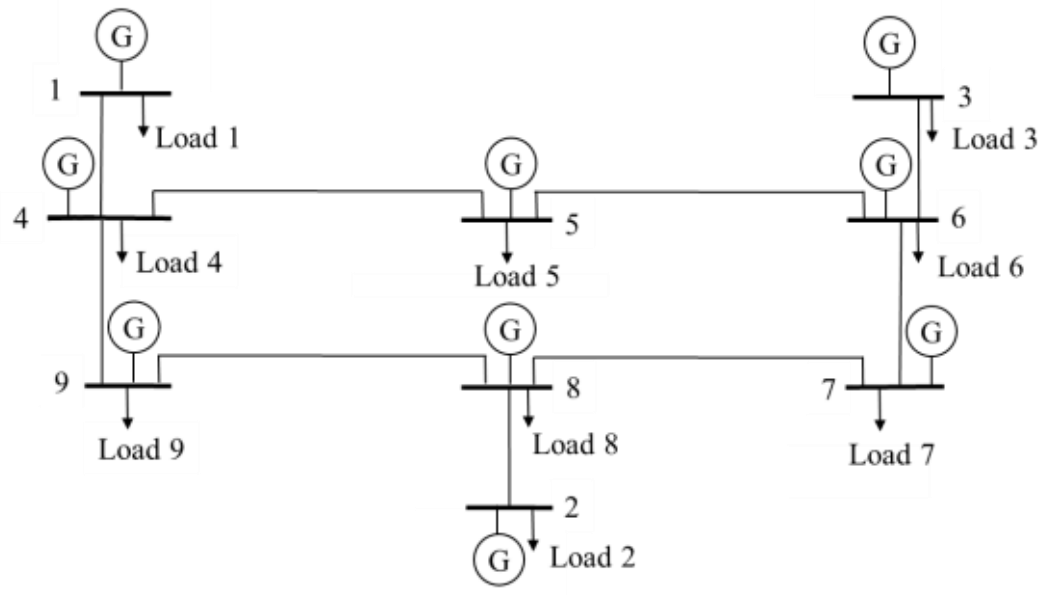

Fig. 2. The IEEE 9 bus system

In this study, it is assumed that seven voltage meters are available. Applying the proposed placing strategy, voltage meters are installed at bus 1, 2, 3, 5, 6, 7, and 9. During a time snapshot, the system has two power sources and one load working, as shown in Fig. 3. The deviation estimation result based on the proposed method is shown in Fig. 4. For comparison, the estimation based on the minimal energy method is shown in Fig. 5. It shows that the proposed method is more accurate compared with the minimal energy estimation.

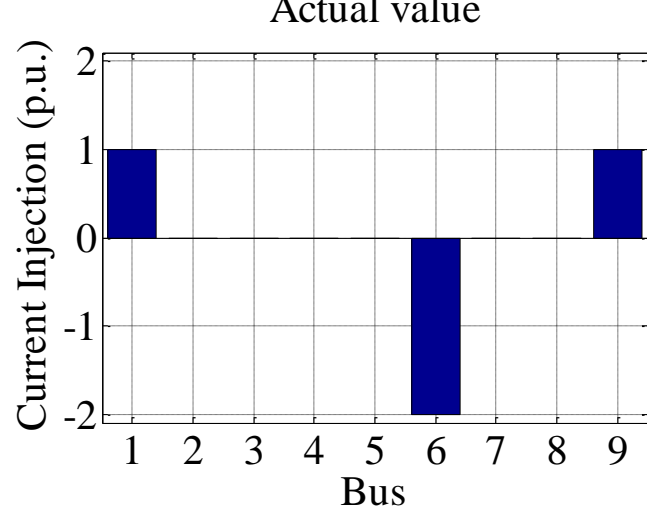

Fig. 3. The actual generator and load current deviation (IEEE-9).

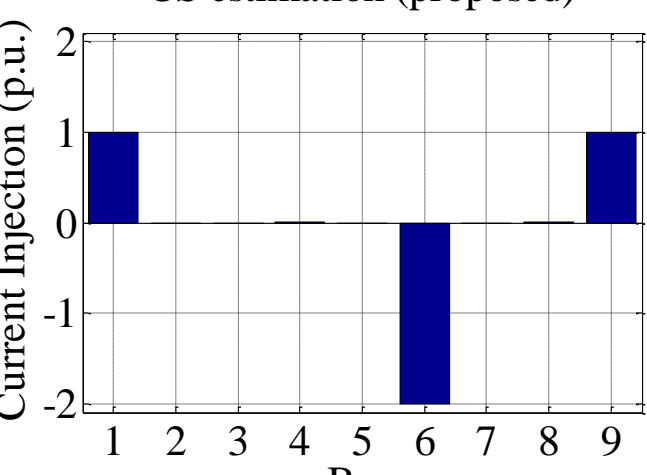

Fig. 4. The estimated generator and load current deviation based on CS and the proposed strategy (IEEE-9).

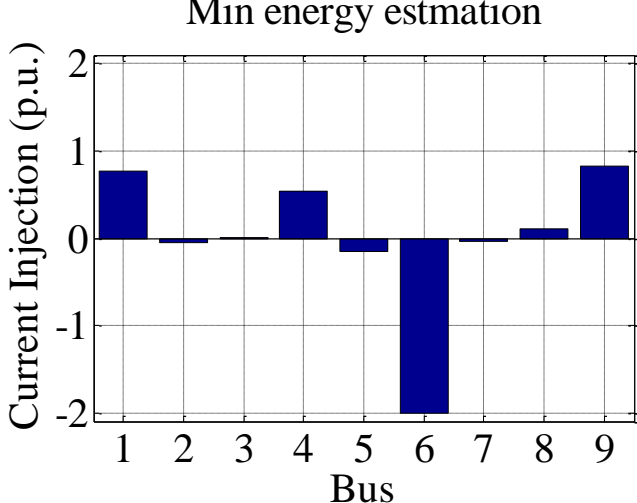

Fig. 5. The estimated generator and load current deviation using minimal energy estimation (IEEE-9).

For comparison, Fig. 6 shows the estimated generator or load injection current deviation based on compressive sensing and random placing of meters. It can be seen that its result quality lies between the proposed method and the minimal energy method. Table I shows the reconstruction ratio of the two methods for meter placement. It can be noted that the proposed method has better performance than the random placement approach, which does not pay special attention to the possible ill-condition feature of the microgrid. This result shows that by placing sensors at locations that resulted to lower coherence in the measurement matrix, the proposed method can effectively achieve the best possible sensor placement and state estimation result.

It is noted that in the real world, there exist many stages of testing before commissioning, and the industry has established and widely-accepted process and systems [29]. The CS-based state estimation and sensor placement proposed in Section II can be plugged into the existing process and systems by selecting the best location for the next available sensor to maximize the system observability and minimize estimation error.

TABLE I: THE SPARSE RECONSTRUCTION RATIO USING THE PROPOSED METER PLACING METHOD AND THE RANDOM PLACING METHOD

| Number of deviated sources and loads to estimate (besides constant resistance load) | Ratio of estimation with < 5% error | | | |
|---|---|---|---|---|
| | Random placement | | Proposed strategy | |
| | 7 meters | 8 meters | 7 meters | 8 meters |
| 1 | 78.5% | 78.8% | 100% | 100% |
| 2 | 51.4% | 61.1% | 82.9% | 91.5% |
| 3 | 42.8% | 51.1% | 75.2% | 87.2% |

*B: IEEE 118 Bus System*

In the IEEE 118 test system, the system has five power sources and five load working during a time snapshot, as shown in Fig. 7. Using the proposed meter placement approach, 90 voltage meters are placed into the system. The estimation results based on the proposed method and the minimal energy method is shown in Fig. 8 and Fig. 9, respectively. Comparison on the results shows that the proposed method is more accurate compared with the minimal energy estimation.

*C. Impact of Meter Errors*

The robustness of the proposed method is studied by incorporating measurement errors before state estimation. It is assumed that all measurements have normal distribution errors with a standard deviation of 0.2%,1%, and 5% in p.u. Changes of the root mean square error of the estimation result with the measurement error is shown in Table II. It can be seen that under all scenarios of sensor accuracy, the proposed method has better accuracy than the minimal energy estimation method. Since compressive sensing reduced the number of sensors below the minimum number of sensors required to make the system fully observable under all conditions, the estimation method is not entirely immune to measurement error. As the estimation is based on under-determined equations formulated by a reduced number of measurements, the accuracy of estimation result is almost linear to the meter accuracy.

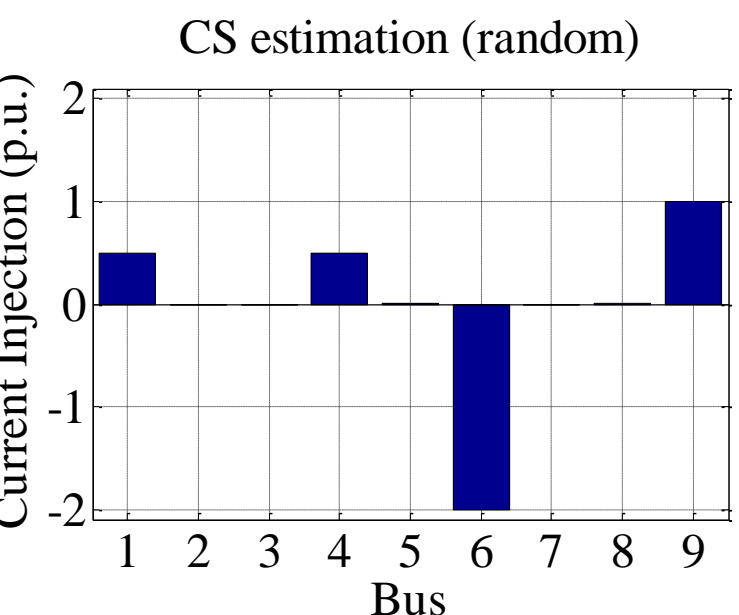

Fig. 6. The estimation result based on CS and random placing of meters.

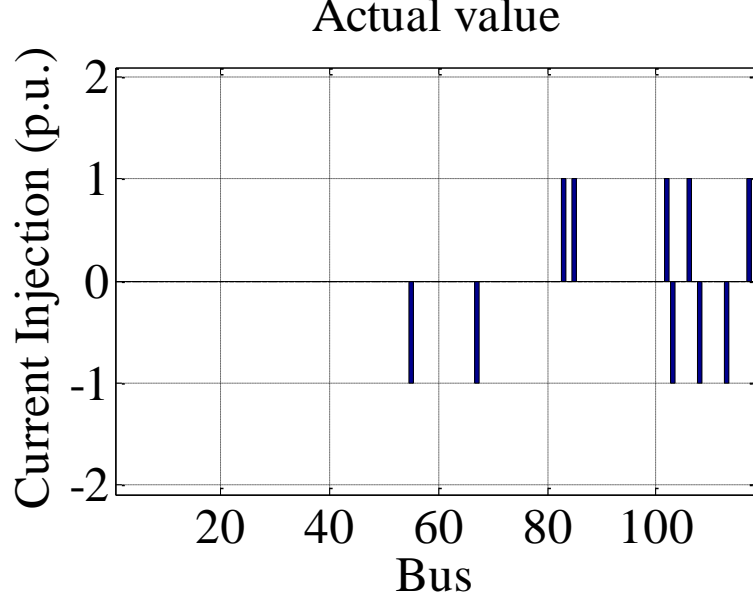

Fig. 7. The actual state deviation of the microgrid (IEEE-118).

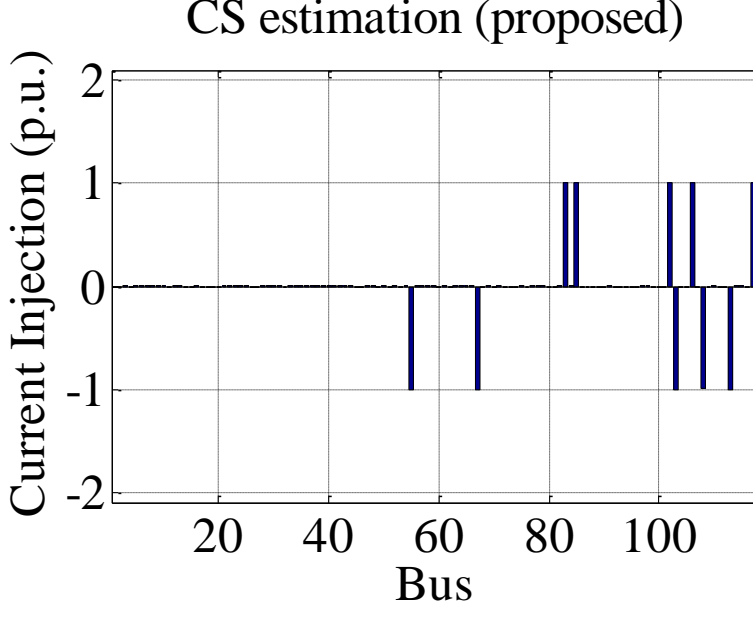

Fig. 8. The state deviation estimation result based on CS and the proposed strategy (IEEE-118).

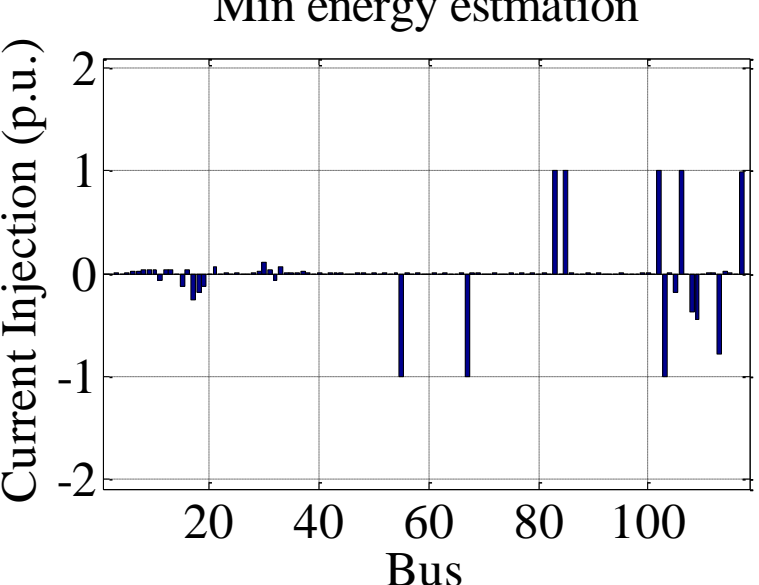

Fig. 9. The deviation estimation result using minimal energy estimation (IEEE-118).

TABLE II: STATE ESTIMATION ACCURACY CHANGES WITH SENSOR ACCURACY

| Meter error standard deviation p.u. | RMSE of estimation result | |
|---|---|---|
| | Compressive sensing | Min energy |
| 0.0% | 0.05% | 8.44% |
| 0.2% | 0.30% | 8.45% |
| 1.0% | 0.96% | 8.49% |
| 5.0% | 4.81% | 9.53% |

### *D. Discussion on Field Implementation*

This study is a proof-of-concept study, which uses MATLAB as the programming language. It should be noted that MATLAB is not a deployment language in protection and automation systems for AC or DC grids. Moreover, compressive sensing is relatively computationally intensive compared with other conventional algorithm for state estimation. More practical programming languages for protection and automation systems, such as C and C++, need to be considered for field implementation [30].

## IV. CONCLUSIONS

The proposed compressive-sensing-based state estimation method can effectively estimate system states using fewer measurements in DC microgrids with relatively fixed operation patterns. The proposed meter placement strategy can minimize the coherence of the measurement matrix to improve estimation performance. The proposed framework shows the potential to achieve more robust state estimation. Future work could be extending the proposed methods to AC and AC/DC hybrid microgrids.